\documentclass[useAMS,usenatbib]{emulateapj} 
\usepackage{epsfig}
\usepackage{lineno}
\usepackage{epsfig}  

\usepackage{amsmath} 
\usepackage{amssymb}
\usepackage{natbib} 
\usepackage{graphicx} 
\usepackage{endnotes}
\usepackage{color}

\shorttitle{First results of AstroSat/LAXPC} \shortauthors{ Yadav et
al.}

\begin{document}

\title{ AstroSat/LAXPC reveals the high energy variability of GRS
1915+105 in the  $\chi$ class} \author{J S Yadav$^1$, Ranjeev
Misra$^2$, Jai Verdhan Chauhan$^1$, P C Agrawal$^3$, H M Antia$^1$,
Mayukh Pahari$^2$, Dhiraj Dedhia$^1$, Tilak Katoch$^1$,
P. Madhwani$^1$, R K Manchanda$^4$, B Paul$^5$, Parag Shah$^1$, C H
Ishwara-Chandra$^6$ } \affil{$^1$ Tata Institute of Fundamental
Research, Homi Bhabha Road, Mumbai, India;
\texttt{jsyadav@tifr.res.in}} \affil{$^2$ Inter-University Centre for
Astronomy and Astrophysics, Pune 411007, India} \affil{$^3$ UM-DAE
Center of Excellence for Basic Sciences, University of Mumbai, Kalina,
Mumbai-400098, India}  \affil{$^4$ University of Mumbai, Kalina,
Mumbai-400098, India} \affil{$^5$ Dept. of Astronomy \& Astrophysics,
Raman Research Institute,  Bengaluru-560080 India} \affil{$^6$
National Center for Radio Astrophysics,  Pune 411007,  India}

\centerline{Accepted for publication in Astrophysical Journal; 23rd August 2016}

\begin{abstract}

We present the first quick look analysis of data from nine {\it
   AstroSat}'s LAXPC observations of GRS 1915+105 during March 2016
   when the source had the characteristics of being in Radio-quiet
   $\chi$ class. We find that a simple empirical model of a disk
   blackbody emission, with Comptonization and a broad Gaussian Iron
   line can fit the time averaged 3--80 keV  spectrum with a
   systematic uncertainty of 1.5\% and a  background flux uncertainty
   of 4\%. A simple deadtime-corrected Poisson noise level spectrum
   matches well with the observed high frequency power spectra  till
   50 kHz and as expected the data show no significant high frequency
   ($> 20$ Hz) features. Energy dependent power spectra reveal a
   strong low frequency (2 -- 8 Hz) Quasi-periodic oscillation (LFQPO)
   and its harmonic along with broad band noise. The QPO frequency
   changes rapidly with flux (nearly 4 Hz in $\sim 5$ hours). With
   increasing QPO frequency, an excess noise component appears
   significantly in the high energy regime ($> 8$ keV).  At the QPO
   frequencies, the time-lag as a function of energy has a
   non-monotonic behavior such that the lags decrease with energy
   till about 15--20 keV and then increase for higher energies.  These
   first look results benchmark the performance of LAXPC  at high
   energies and confirms that its data can be used for more
   sophisticated analysis such as flux or frequency-resolved
   spectro-timing studies.

\end{abstract}


\section{Introduction}\label{intro}

The extraordinary micro-quasar black hole system, GRS 1915+105 was the
primary subject of a large number of Rossi X-ray Timing Explorer (RXTE) observations that has revealed its complex
spectral and timing behavior which needs to be classified into more
than twelve distinct variability classes \citep{be00}. It is the first
Galactic black hole X-ray binary where superluminal jet has been
discovered \citep{fe99}. Extensive study revealed variations in its
radio flux (at 15.2 GHz) from few  to few hundreds of milliJy in
different variability classes \citep{mu01} as well as in the same
variability class \citep{pa13b}. This source has been extensively
studied to understand the disk-jet connection in X-ray binaries
\citep{mi98,ya01,fe04,ya06}.

\begin{figure}
\centering
\includegraphics[width=0.34\textwidth,angle=-90]{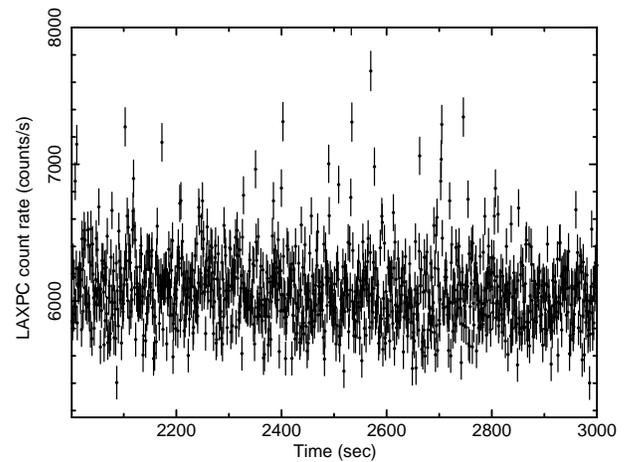}
\caption{A typical 1 ksec lightcurve of the $\chi$ class in the energy
range 3.0--80.0 keV is shown where count rate from all three LAXPC
detectors --- {\tt LAXPC10}, {\tt LAXPC20} and {\tt LAXPC30} are
combined.}
\label{lightcurve}
\end{figure}

 During some of these classes the source shows large amplitude
 variability in flux and spectral shape on time-scales of minutes to
 hours \citep{ta97,paul98,ya99,pa13a,pa13c}, while in others the
 source is relatively steady. Even when the source is relatively quiet
 on minute time-scales, detailed analysis often show strong coherent
 quasi-periodic oscillations (QPOs) with frequencies of the order of a
 few Hz \citep{paul97,mo97,mu99,pa13b}. Interestingly, one of the
 promising model for these oscillations is that the variation is
 produced by the inner disk globally precessing due to the
 Lens-Thirring effect around a spinning black hole
 \citep{st98,in09}. Evidence for such a behavior is inferred from the
 time-averaged spectrum and the variation of the QPO properties with
 intensity and corresponding spectral parameters \citep{in15}. Perhaps
 more importantly, energy dependent QPO properties such as fractional
 rms and time-lag between different energy bins provide crucial
 information regarding such phenomena \citep{re00,mu01,qu10,pa13b}.

\begin{table*}
 \centering
 \caption{Comparison of approximate LAXPC characteristics with those of
RXTE/PCA and NuSTAR}
\begin{center}
\scalebox{0.95}{%
\begin{tabular}{|l|l|l|l|}
\hline  Instrument & AstroSat/LAXPC (i) & RXTE/PCA (ii) \& (iii) & NuSTAR (iv) $\&$ (v) \\ 
\hline
Detector type & Non--imaging & Non--imaging & Imaging \\ 
\hline 
Energy range (keV) & 3--80 & 2--60 & 3--78  \\ 
\hline 
Dead time & $\sim$42 $\mu$s & $\sim$10 $\mu$s & 1--2 ms \\ 
\hline 
Energy Resolution &  &  & \\
 At 6 keV &  $\sim$20 \%  & $\sim$18 \%  & $\sim$6.67 \%  \\
At 60 keV &  $\sim$10 \%  & -- & $\sim$1.5 \%  \\
\hline 
Effective area (cm$^{2}$) $^{*}$  &  &  &  \\ 
At 10 keV & $\sim$6100 & $\sim$7000 & $\sim$800 \\ 
At 30 keV & $\sim$4500 & $\sim$1000 & $\sim$300 \\ 
At 50 keV & $\sim$5100 & $\sim$750 & $<$200 \\ 
\hline
\end{tabular}}
\tablecomments {$^{*}$ The LAXPC effective area quoted here is based on simultaneous fit to LAXPC and NuSTAR   data. More reliable estimate will be available after simultaneous observations with several observatories are performed. References: [i]\citet{ya16} , [ii]\citet{za93} , [iii]\citet{ja06}, [iv]\citet{ha13}, [v]\citet{ba15} }
\end{center}
\label{obs1}
\end{table*}

Most of the combined  spectral and timing analysis of such sources
like GRS 1915+105, have been done using the {\it RXTE} Proportional
Counter Array (PCA) data,  which despite the wealth of information
they provided, were limited due to several reasons. Due to telemetry
constraints, the {\it RXTE} data of bright sources on many occasions
were binned in channels on-board, leaving only a few broad energy
bands for which analysis could be undertaken. The effective area of
the proportional counter units (PCUs) decreased rapidly beyond 30 keV
\citep{ja06} and hence the statistics for high energy photons were
usually not sufficient to do any detailed timing analysis.  Despite
efforts to improve the calibration \citep{sh12}, the spectral analysis
was typically limited to the 3--30 keV band and hence one had to rely
on rare simultaneous observations of the source from other satellites
to obtain broadband spectral data.

{\it AstroSat}, the first Indian astronomical satellite which was
successfully launched on September 28, 2015 has five scientific
instruments on board \citep{ag06,si14}. One of the primary instruments is the Large Area X-ray Proportional counters
(LAXPC) which consists of three identical proportional counter units
adding up to an effective area of $\sim 6000$ cm$^2$ at
15 keV \citep{ya16,ag16}. After launch, LAXPC went through a series
of performance  verification tests where it observed different blank
sky coordinates, standard sources and sources of scientific
interest. These tests have allowed for the calibration and estimation
of the instrument spectral response, background characteristic and
timing capabilities which will be  presented in detail elsewhere.
Here we provide a first  estimate of the characteristics of the instrument
and compare them with those of {\it RXTE}/PCA and {\it NuSTAR} in 
Table \ref{obs1}. While {\it NuSTAR} is an imaging instrument, the
LAXPC provides several advantages  as compared to the {\it
RXTE}/PCA. The effective area of the LAXPC at energies greater than 30
keV is significantly larger than the PCA (Table \ref{obs1}). The complete event  data obtained allows for energy
dependent analysis of any choice of energy bins. The other co-pointing
instruments on board {\it AstroSat} especially the Soft X-ray
telescope (SXT) can provide critical spectral coverage below 3 keV.

\begin{table*}
 \centering
 \caption{LAXPC observation details of GRS 1915+105}
\begin{center}
\scalebox{0.95}{%
\begin{tabular}{|l|l|l|l|l|l|}
\hline  Orbit number & Date & Start time  & Effective on-source &
average source count & QPO frequency  \\ & (dd-mm-yyyy) & (hh:mm:ss) &
exposure (sec) & rate of {\tt LAXPC10} & (Hz) \\ \hline  2359 &
05-03-2016 & 09:54:19.24 & 1728.0 & $3697 \pm 30$ & $5.72 \pm 0.04$ \\
\hline 2360 & 05-03-2016 & 11:33:47.19 & 721.0 & $3637 \pm 30$ & $6.55
\pm 0.05$ \\ \hline 2361 & 05-03-2016 & 13:12:31.66 & 1439.0 & $2855
\pm 30$ & $4.53 \pm 0.03$ \\ \hline 2362 & 05-03-2016 & 14:54:35.59 &
2165.0 & $2400 \pm 26$ & $3.49 \pm 0.03$ \\ \hline 2363 & 05-03-2016 &
16:23:03.34 & 2292.0 & $2209 \pm 11$ & $2.53 \pm 0.02$ \\ \hline 2364
& 05-03-2016 & 18:08:11.91 & 2716.0 & $2219 \pm 12$ & $2.79 \pm 0.03$
\\ \hline 2365 & 05-03-2016 & 19:48:03.92 & 3148.0 & $2313 \pm 19$ &
$3.18 \pm 0.03$ \\ \hline 2367 & 05-03-2016 & 23:35:54.71 & 1913.0 &
$3022 \pm 24$ & $4.61 \pm 0.03$ \\ \hline 2368 & 06-03-2016 &
01:34:36.34 & 2873.0 & $3334 \pm 27$ & $4.94 \pm 0.04$ \\ \hline
\end{tabular}}
\end{center}
\label{obs}
\end{table*}

As a part of the performance verification, during  5--7 March 2016,
{\it AstroSat} observed the black hole system GRS 1915+105 for many orbits and for nine of them the flux was relatively
steady on time-scales of minutes suggestive that they belong to
Radio-quiet $\chi$ class. GRS 1915+105 moves to this class from
$\theta$ class and at the end of $\chi$ class, the source again
returned to the $\theta$ class. Therefore, GRS 1915+105 remains for
$\sim 14$ hrs in the Steep Power Law (SPL) state which is a subclass
of $\chi$ class. X-ray activity in GRS 1915+105 have been
classified in twelve different classes. $\chi$ class is characterized
by hard spectrum with low frequency QPO while $\theta$ class is a
flaring X-ray Class associated with transient radio jets every 20 -- 30
minutes \citep{be00}.

In this work, we present the first look spectral and timing analysis
of these observations, with particular emphasis on the high energy
behavior.

\section{Spectral Analysis}

\begin{figure}
\centering
\includegraphics[width=0.32\textwidth,angle=-90]{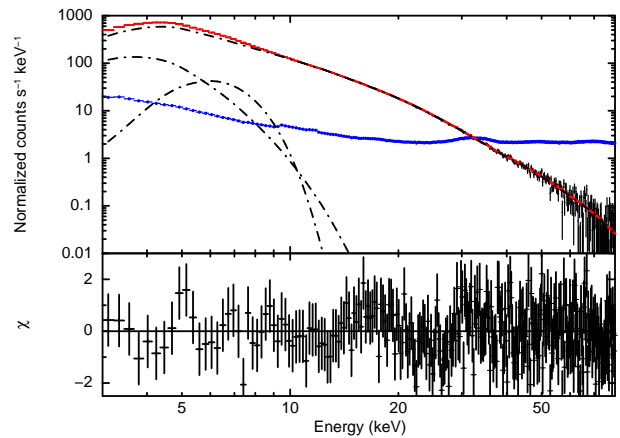}
\caption{The top panel shows counts spectrum as observed from {\tt
LAXPC10} for the orbit number 2363, fitted with a model consisting of
a disk black body, thermal Comptonization and a broad Iron
line. Resultant model spectra is shown in red line while the thick
blue line represents the expected background level. The bottom panel
shows the residuals which may be due to systematic errors in the
response function. The spectrum provides a good fit of
$\chi^2/\mathrm{dof} = 278/284$ when a 1.5\% systematics are added and
the background uncertainty is taken to be 4\% . }
\label{spec}
\end{figure}

In this work we present analysis of nine LAXPC observations of GRS
1915+105 as listed in Table \ref{obs}. Each of them were obtained from
a single orbit with effective exposure of $\sim$ 1--3 ksec each. A
typical 1 ksec lightcurve observed during orbit no. 2363 is shown in
Figure \ref{lightcurve} where count rate from all three LAXPC units
are combined. We begin with the time averaged spectral analysis of
these observations. While the complete detailed description of LAXPC
performance and calibration will be presented elsewhere \citep{ya16},
here we highlight some of the basic points. The response function for
each of the LAXPC units was computed using {\tt GEANT4} simulations.

Most of the background is coming from cosmic diffused X-ray background. The background for each of the unit was modeled as a 
function of the latitude and longitude of the satellite \citep{an16}. There are differences between the observed background and
the model at a level of about 4\% which  is not understood so far.
In this work, the timing analysis is for frequencies much larger than 0.01 Hz and the background
is stable on these timescales.  We have checked that the power spectra 
for Crab and the background show no spurious signals at these frequencies.
 The source spectra,
lightcurves and background spectra were extracted using software which
will  become part of the standard pipeline for the LAXPC. The spectral
fitting was done using the {\tt XSPEC v 12.8.1g} spectral fitting
package.

The LAXPC consists of three identical proportional counter detectors
units which are named as {\tt LAXPC10}, {\tt LAXPC20} and {\tt
LAXPC30}.  We checked that joint spectral fitting of {\tt LAXPC10} and
simultaneous {\it NuSTAR} observations of Crab produced the expected
best fit power-law index of $2.08\pm0.01$ with the
normalization constant of LAXPC relative to {\it NuSTAR}
giving the effective area quoted earlier. For the fitting a 4\%
uncertainty was added to the background (which needs to be dead time
corrected for bright sources) and a 1.5\% overall systematic is added
to the spectrum to account for the uncertainty in the response. We use
these values throughout this work.
The complete energy spectrum of
the LAXPC 10 were represented by 512 pha bins of which 292 of them
lie within the energy range of 3--80 keV. This gives an average
energy bin size of $\sim 0.3$ keV which is approximately 
a fourth of the approximate energy resolution of LAXPC at 6 Kev i.e.
$\sim 1.2$ keV. However at higher energies the spectra would
be oversampled and more detailed later work should rebin the 
spectra based on the energy resolution, once the response of the
detector is better understood and modeled.

\begin{figure}
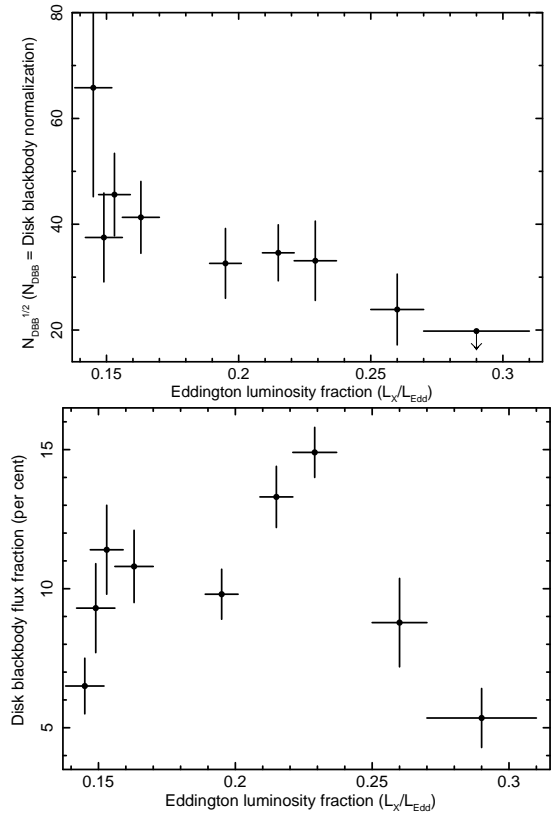

\centering
\includegraphics[width=0.3\textwidth,angle=-90]{lum-norm.ps}
\centering
\includegraphics[width=0.3\textwidth,angle=-90]{lum-diskfluxfrac.ps}
\caption{The square root of disk blackbody normalization (Top panel)
and the percentage fraction of the disk to the total flux as a
function of Eddington ratio (bottom panel) are shown for different
observations.For computing the Eddington ratio we used a black hole
mass of $ 12.5 M\odot$ and a distance to the source $D = 9$ kpc
\citep{re14}.}
\label{lumvar}
\end{figure}

Figure \ref{spec} shows the counts spectrum of {\tt LAXPC10} for the
observation of orbit no.~2363 which was fitted by a model consisting
of a disk black body Component (XSPEC model {\tt DISKBB}), a thermal
Comptonized component (XSPEC model {\tt NTHCOMP}) and a broad Gaussian
feature (XSPEC model {\tt GAUSSIAN}) to represent the Iron line. The
disk blackbody component was taken to be the seed photon source for
the Comptonization. All spectra are modified by the
presence of neutral Hydrogen absorption which has been taken care by
using TBabs model \citep{wi00}.  
GRS 1915+105 has high intrinsic
absorption with equivalent neutral column density of the order of few times
$ 10^{22}$ cm$^{-2}$ \citep[e.g.][]{mc06}. The LAXPC data is sensitive
only beyond 3 keV making it difficult to constrain the column density
and hence it was fixed to a value of $ 6 \times 10^{22}$ cm$^{-2}$.
The thick blue line represents
the expected background count rate for the observation. The bottom
panel shows strong residuals which may be due to uncertainties in the
response function. Indeed, if an overall systematic of 1.5\% is added (with 4\%
uncertainty in background),
the fitting gives an acceptable $\chi^2/\mathrm{dof} =
278/284$ in place of $\chi^2/\mathrm{dof} = 1048/284$ (without the
systematic). Some of the parameters obtained were the disk
temperature $kT = 0.88^{+0.05}_{-0.08}$ keV, the Comptonization photon
index $\Gamma = 2.57^{+0.03}_{-0.02}$ and electron temperature $kT_e =
16.8^{+1.05}_{-1.28}$ keV. Similar fits were obtained for all nine
observations. While detailed tabulation of the best fit spectral
parameters using different spectral models will be shown elsewhere,
here we highlight some of the generic results.
The normalization of
the disk blackbody emission seems to decrease slightly with luminosity
(top panel of Figure \ref{lumvar}) but
the large errorbars do not allow for concrete statements to be made.
For the mass and
distance assumed for this source, the typical value of the normalization 
would imply an inner radius of
$\sim 6 (f/1.7)^2 GM/c^2$  for a color factor $f \sim 1.7$
\citep{sh95}, suggesting that disk extends to be truncated
for a fast spinning blackhole \citep{mc06}. Moreover, for the nine observations, the temperature of the
Comptonizing medium was found to be around $\sim15$ keV.  The fraction
of the disk component flux to the total does not seem to
show any correlation as shown in the bottom panel of Figure
\ref{lumvar}. However, more detailed investigations taking into
account more physical models like gravitationally smeared disk and
Iron line emission (instead of the simple {\tt DISKBB} model and broad
Gaussian used here) need to be undertaken before any inferences can be
made. We defer such analysis to the future when we expect to have
significantly better instrument response and background modeling than
the currently available version.

\section{Variability Analysis}
\subsection{High frequency variability}
The high telemetry of {\it AstroSat} allows for LAXPC data to be
obtained in event mode for high count rate sources at the instrument
time resolution. This provides an unprecedented opportunity to study
the rapid variability of bright sources.

\begin{figure}
\centering
\includegraphics[width=0.3\textwidth,angle=-90]{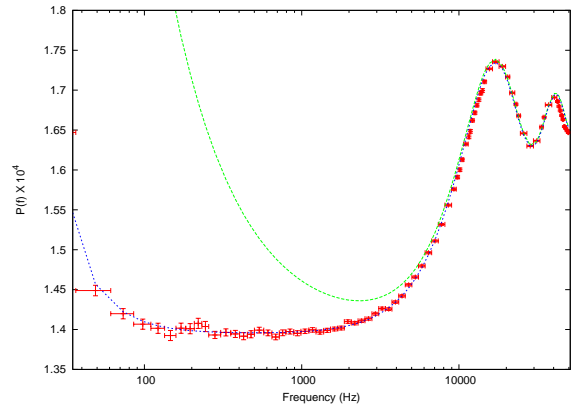}
\caption{The high frequency rebinned power spectra for a 2.3 ksec
observation of  GRS 1915+105 in the $\chi$ class. The power spectra
matches well with the predicted Poisson noise level with a dead time
$\tau = 42.3$ microseconds and a low frequency power-law
component. The green line shows the expected peak power of a QPO with
quality factor $Q = 4$ and rms of 5\%. LAXPC would have detected a QPO
of such a strength easily till 3000 Hz.}
\label{power_all}
\end{figure}

As an example to demonstrate the capability of LAXPC to detect high
frequency variability, we consider just one observation of effective
exposure $\sim 2.3$ ksec, i.e., orbit no.~2363. Using the event mode
data at the time resolution of 10 microseconds, the power spectrum up
to a Nyquist frequency of 50 kHz was computed for counts in the energy
range 3.0--80.0 keV. Data from all three units were combined. The
light curve was divided into  segments of 2048 bins corresponding to
0.02048 seconds. The total exposure consists of 97655 segments. The
power spectrum generated at 1024 frequency points was rebinned and
shown in Figure \ref{power_all}. The rise in the power spectrum at low
frequencies ($< 60$ Hz) is due to low frequency variability of the
source while the structure at high frequencies is due to the
characteristic effect of dead time. The expected dead time corrected
Poisson level power for a system which is non-paralyzable is given by
\citet{zh95}

\begin{eqnarray}
P_N(f) = \; & \frac{1}{R_{BO}^2}\left[2 +
4\frac{<R_{B}>}{R_T}\hspace{1.2in} \nonumber \right.\\ & \left. \times
\frac{-1+\cos(2\pi f\tau_d) -( 2\pi f\tau) \sin(2\pi f\tau_d)}{2 +
(2\pi f\tau)^2 - 2\cos(2\pi f\tau_d) + 2\pi f\tau \sin(2\pi
f\tau_d)}\right]
\end{eqnarray}
Here, $\tau_d$ is the dead time and $R_{BO}$ is the observed rate in
the energy band being considered. $R_T$ is the dead-time corrected
observed count rate for the detector, i.e.  $R_T =
R_{TO}/(1-R_{TO}\tau_d)$ where $R_{TO}$ is the observed total count
rate and $\tau = 1/R_{T}$.  $<R_{B}>$ is the average dead time
corrected count rate per detector i.e. $<R_{B}> =
R_{BO}(1+R_T\tau_d)/N_{pcu}$ with $N_{pcu}$ being the number of LAXPC
units being considered.

\begin{figure}
\centering
\includegraphics[width=0.3\textwidth,angle=-90]{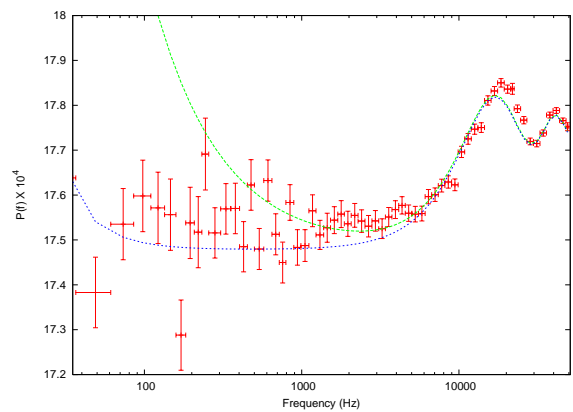}
\caption{Same as Figure \ref{power_all}, except that the power
spectrum and the expected dead time corrected Poisson level is
computed for counts in the high energy 20--80 keV band. While there
are hints of possible features they are not significant. }
\label{power_high}
\end{figure}

For the low frequency variability we consider a power-law $\propto
f^{-2}$ and hence the total function is taken to be $P (f) =
A(f/10\,\mathrm{Hz})^{-2} + P_N(f)$ and we fit the unbinned observed
power spectrum consisting of 1024 frequency bins. The mean effective
dead time as measured on the ground was around 50 microseconds but it
has variations depending on the kind of event detected and hence is
not a constant as assumed in the above equation. Hence, we  attempt to
obtain an average value of dead-time by fitting the parameters, $A$
and $\tau_d$ and keeping the rest of the parameters fixed at their
observed values, i.e., $R_{BO} = 5999.7$ and $R_{TO} = 6236.4$
counts/s. The fit gave a $\chi^2/dof = 2349.8/1022$ or a reduced
$\chi^2_{red} = 2.3$. The best fit average dead time was found to be
$\tau_d = 42.3$ microseconds. The best fit curve is shown
as a dotted blue line in Figure \ref{power_all}. Thus, considering
all the uncertainties, the expected dead time corrected power spectrum
is remarkably close to the observed one. In fact, a $0.5$ percent
systematic uncertainty added in quadrature to the power spectrum
yields a reduced $\chi^2_{red} = 0.7$, showing the level at which the
power spectrum can be trusted in these high frequencies. The
sensitivity of the instrument can be further seen by the dotted line in Figure \ref{power_all}, which represents the
expected peak of the power spectrum for a Lorentzian shaped
Quasi-periodic Oscillation (QPO), with a Quality factor $Q = 4$ and
rms of 5\%, i.e.,  $P_{QPO} =\frac{2 Q}{\pi
f}(\mathrm{rms})^2$. Thus, a QPO of this strength and
quality factor would have easily been detected by LAXPC for a source
like GRS 1915+105  in a 3 ksec observation if its frequency was less
than 3000 Hz.

Since high frequency features when seen in black hole 
systems are stronger at higher energies and sometimes can only be 
detected at  high energies
\citep{cu00,ho03}, we verify LAXPC capability to detect such QPOs in
the 20--80 keV band. In this band for the observation under
consideration the count rate observed was 396 counts/sec. Figure
\ref{power_high} shows the rebinned power spectrum and the expected
Poisson noise level at this energy band. While some features can be
seen, they are not significant in the sense that a fit to the power
spectrum using just the noise level and low frequency component, gives
a reduced $\chi^2_{red} = 1.05$. The expected peak power for a QPO
with the quality factor Q = 4 and rms 5\% (green line in Figure
\ref{power_high}) shows that such a QPO would have been only
marginally detected by LAXPC for this observation.

\subsection{Hardness-Intensity and colour-colour variations}

\begin{figure}
\centering
\includegraphics[width=0.3\textwidth,angle=-90]{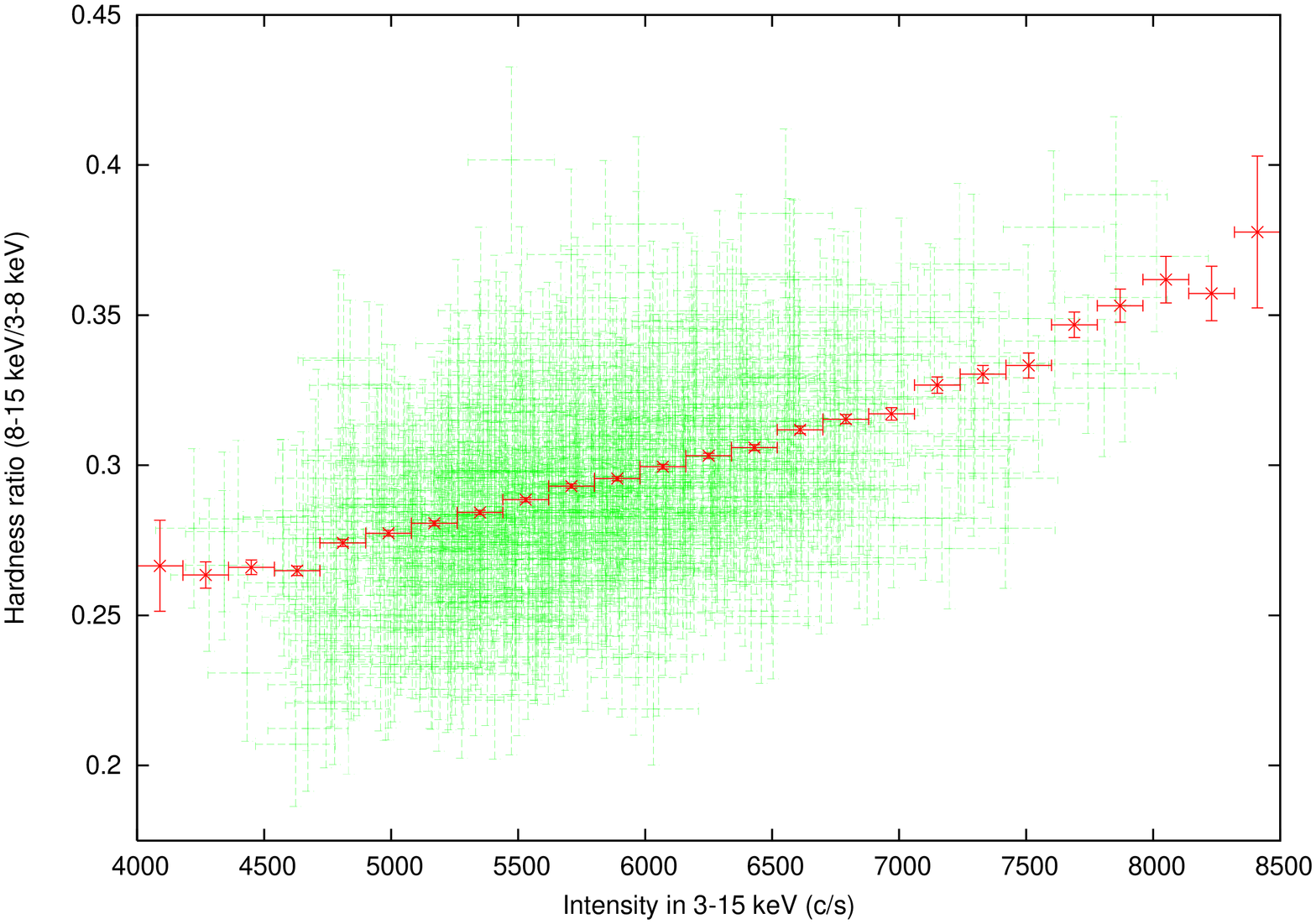}
\centering
\includegraphics[width=0.3\textwidth,angle=-90]{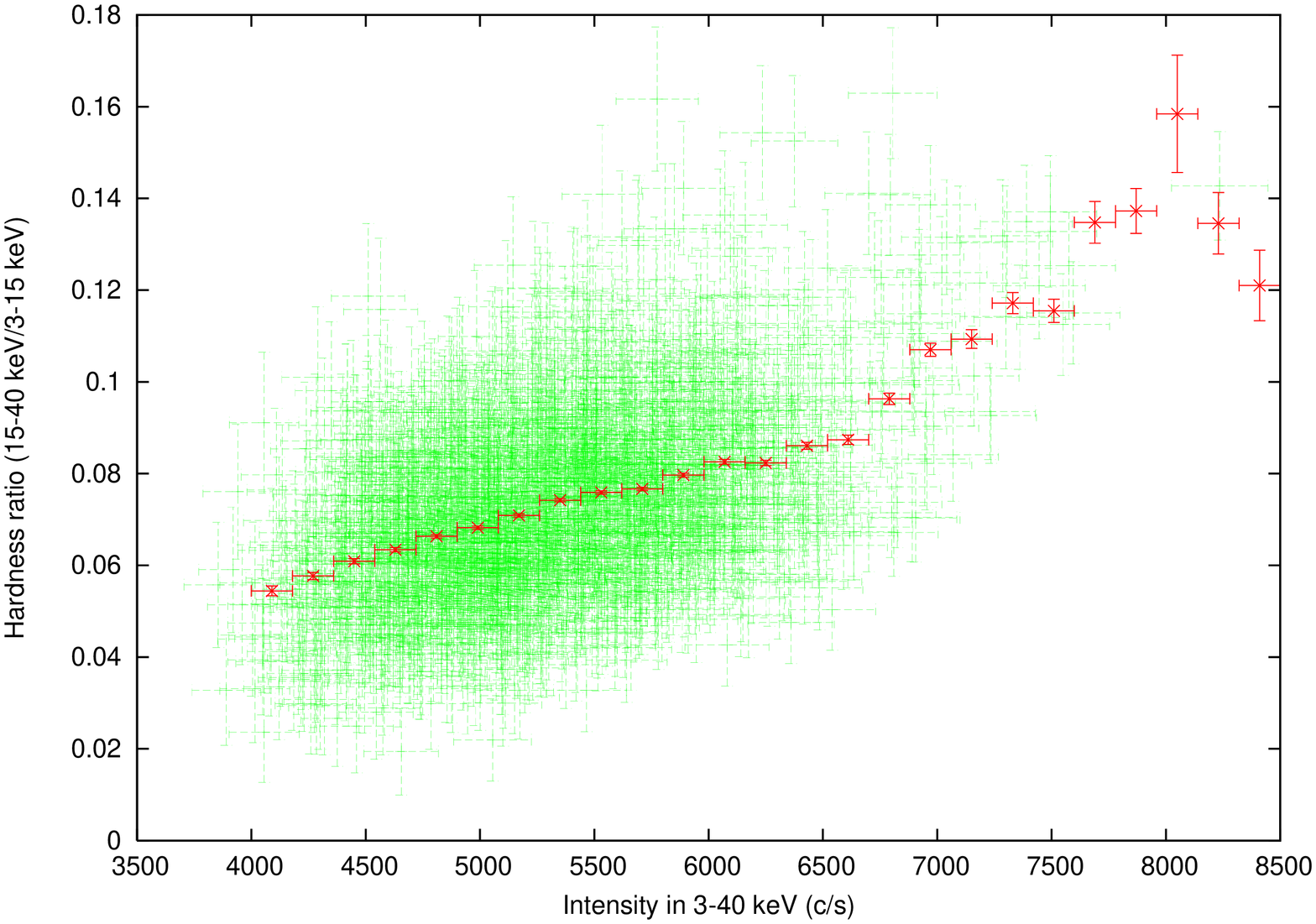}
\caption{Hardness Intensity plot for 2.3 ksec observation of GRS
1915+105 in 0.25 second time bin. About 20\% of the total points are
shown for clarity. In the top panel, the hardness is defined as the
count rate ratio  between 8--15 and 3--8 keV bands while the intensity
is the total count rate in 3--15 keV band. In the bottom panel, the
ratio is similarly defined for 3--15 and 15--40 keV and the total
intensity in the 3--40 keV band. The red points are the average values
at a given intensity. The source clearly shows increasing hardness
with  intensity in both soft and hard bands.}
\label{hardness}
\end{figure}

\begin{figure*}
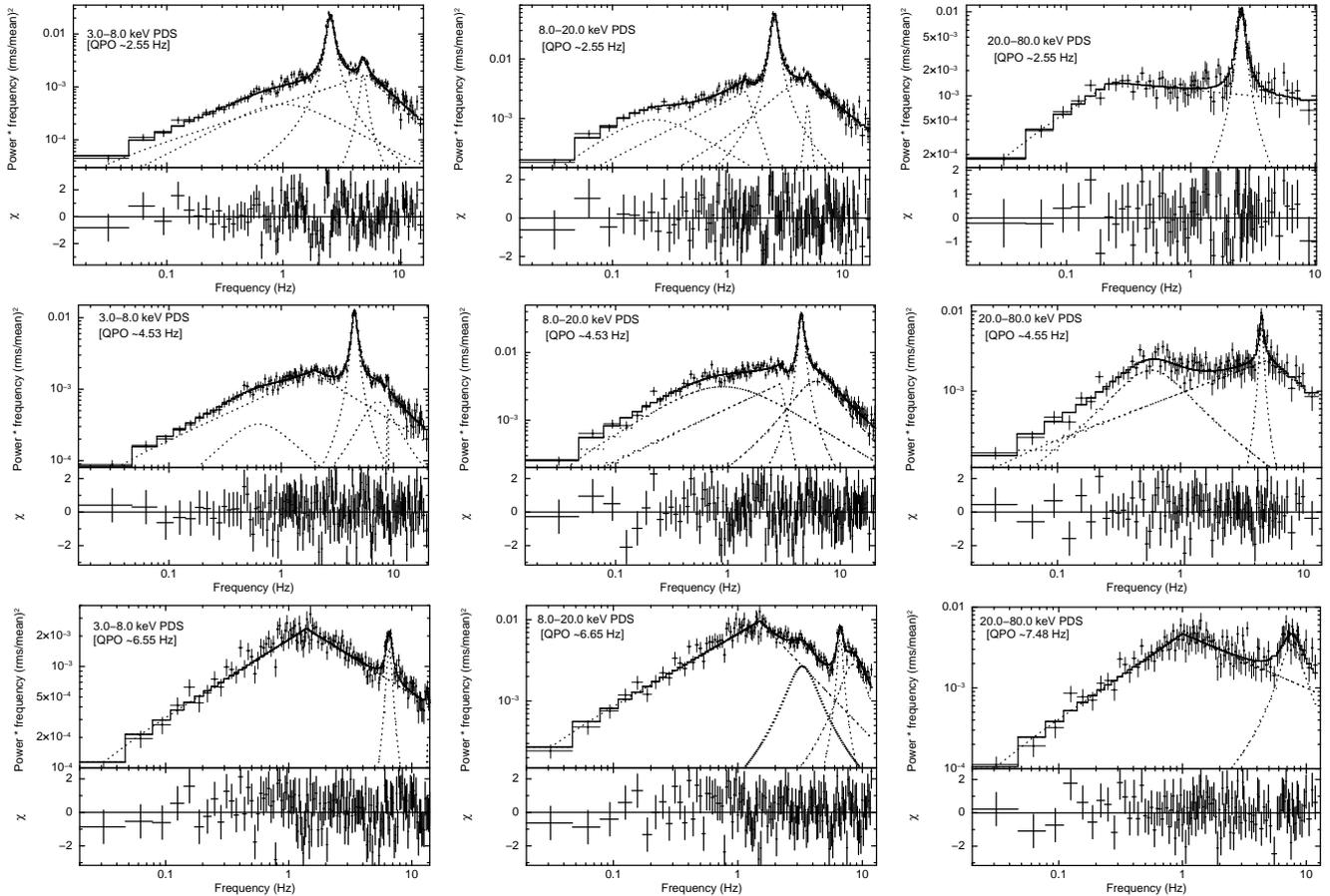

\centering \includegraphics[width=0.22\textwidth,angle=-90]{fig4a.ps}
\centering \includegraphics[width=0.22\textwidth,angle=-90]{fig4b.ps}
\centering \includegraphics[width=0.22\textwidth,angle=-90]{fig4c.ps}
\centering \includegraphics[width=0.22\textwidth,angle=-90]{fig4d.ps}
\centering \includegraphics[width=0.22\textwidth,angle=-90]{fig4e.ps}
\centering \includegraphics[width=0.22\textwidth,angle=-90]{fig4f.ps}
\centering \includegraphics[width=0.22\textwidth,angle=-90]{fig4g.ps}
\centering \includegraphics[width=0.22\textwidth,angle=-90]{fig4h.ps}
\centering \includegraphics[width=0.22\textwidth,angle=-90]{fig4i.ps}
\caption{Power density spectra at three different energy bands
3.0--8.0 keV (first column), 8.0--20.0 keV (second column) and
20.0--80.0 keV (third column) are shown for three observations when
strong QPOs are detected at $\sim 2.55$ Hz (top rows), $\sim 4.53$ Hz
(middle rows) and $\sim 6.55$ Hz (bottom rows) respectively. Due to
observed break in noise continuum, broken powerlaw model is used to
fit the noise component while Lorentzians are used to fit QPO and
harmonic features. It may be noted that a significant, excess noise
component (modeled with broad Lorentzian) appears in the PDS with
higher QPO frequencies at the energy $> 8.0$ keV. Such features were
not detected by {\it RXTE} due to its highly reduced efficiency in
8.0--20.0 keV energy band.}
\label{power}
\end{figure*}

\begin{figure}
\centering \includegraphics[width=0.27\textwidth,angle=-90]{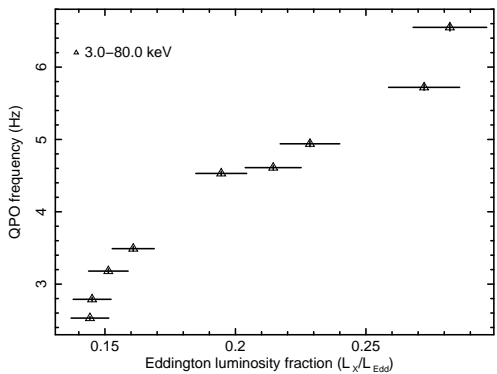}
\caption{The primary QPO frequency is shown as a function of the ratio
of source luminosity to Eddington luminosity. QPO frequency is
observed to be monotonically increasing with the source luminosity.}
\label{QPOEdd}
\end{figure}

Having established the expected result that the  source does not show
any strong high frequency phenomena, we turn our attention to the
behavior of the source in low frequencies i.e., $< 10$ Hz. To begin,
we compute hardness intensity and color-color plots to understand the
overall spectral variability of the source. Light curves were
extracted from all three units in different energy bands and the dead
time corrected background in that band (i.e., $R_{BC} =
R_B/(1+R_T\tau_d) $) was subtracted.

The top panel of Figure \ref{hardness} shows the hardness ratio
between the 8--15 and 3--8 keV band versus the intensity in the 3--15
keV with a time bin of 0.25 secs. The source clearly hardens as it
gets brighter which is better illustrated when the points
corresponding to an intensity bins are averaged (solid line in the
Figure). More remarkable is that LAXPC allows a hardness intensity
plot to be made in a broader energy ranges of 3--15 and 15--40 keV
bands as shown in the bottom panel of Figure \ref{hardness}. The
behavior of the source is found to be similar to the softer
bands. That such an analysis can be undertaken shows that LAXPC data
is well suited for more sophisticated analysis such as flux or even
frequency resolved spectroscopy which will be undertaken in a future
work.

\subsection{Energy dependent power spectra}

\begin{figure*}
\centering \includegraphics[width=0.22\textwidth,angle=-90]{fig5a.ps}
\centering \includegraphics[width=0.22\textwidth,angle=-90]{fig5b.ps}
\centering \includegraphics[width=0.22\textwidth,angle=-90]{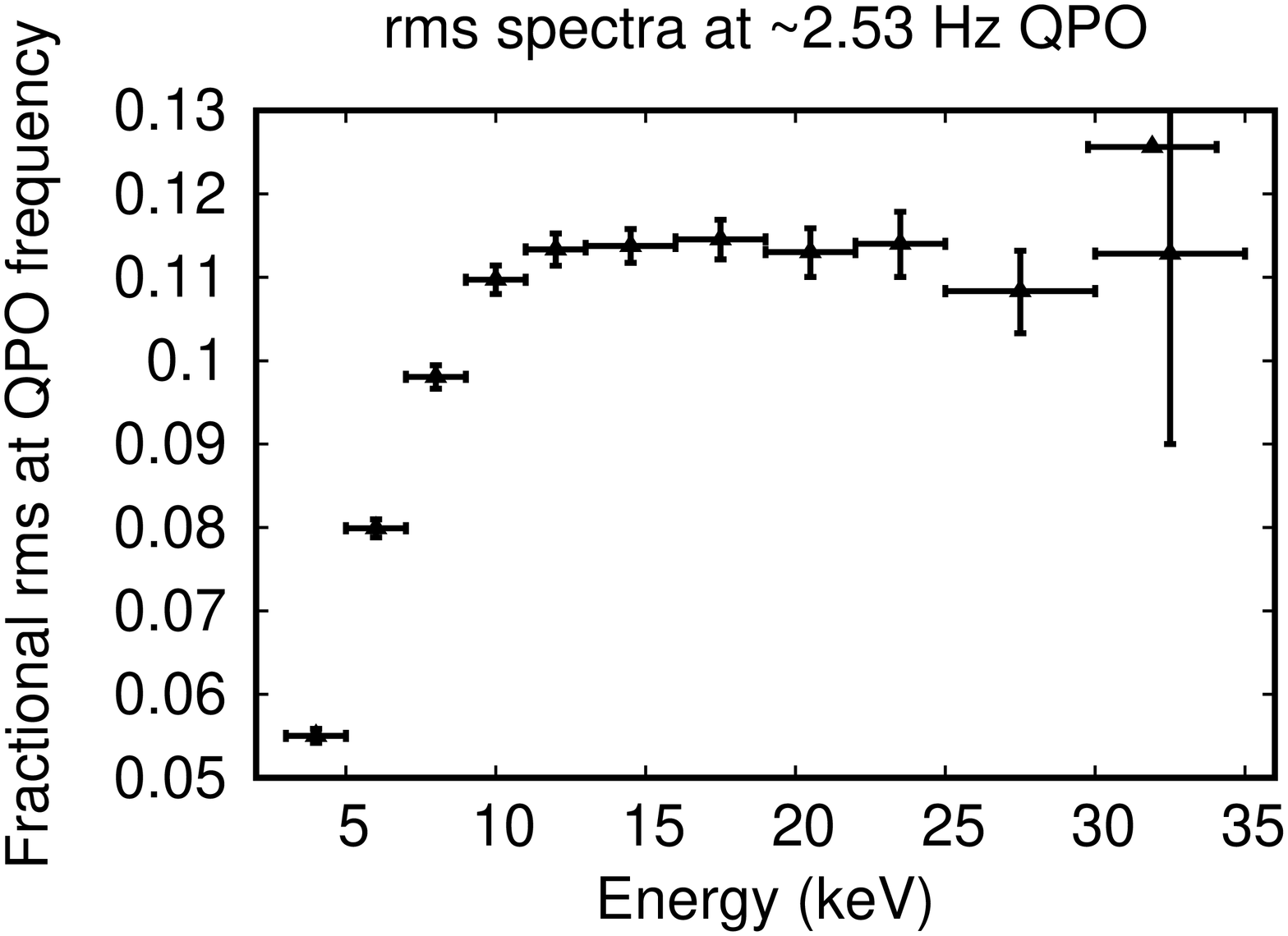}
\centering \includegraphics[width=0.22\textwidth,angle=-90]{fig5d.ps}
\centering \includegraphics[width=0.22\textwidth,angle=-90]{fig5e.ps}
\centering \includegraphics[width=0.22\textwidth,angle=-90]{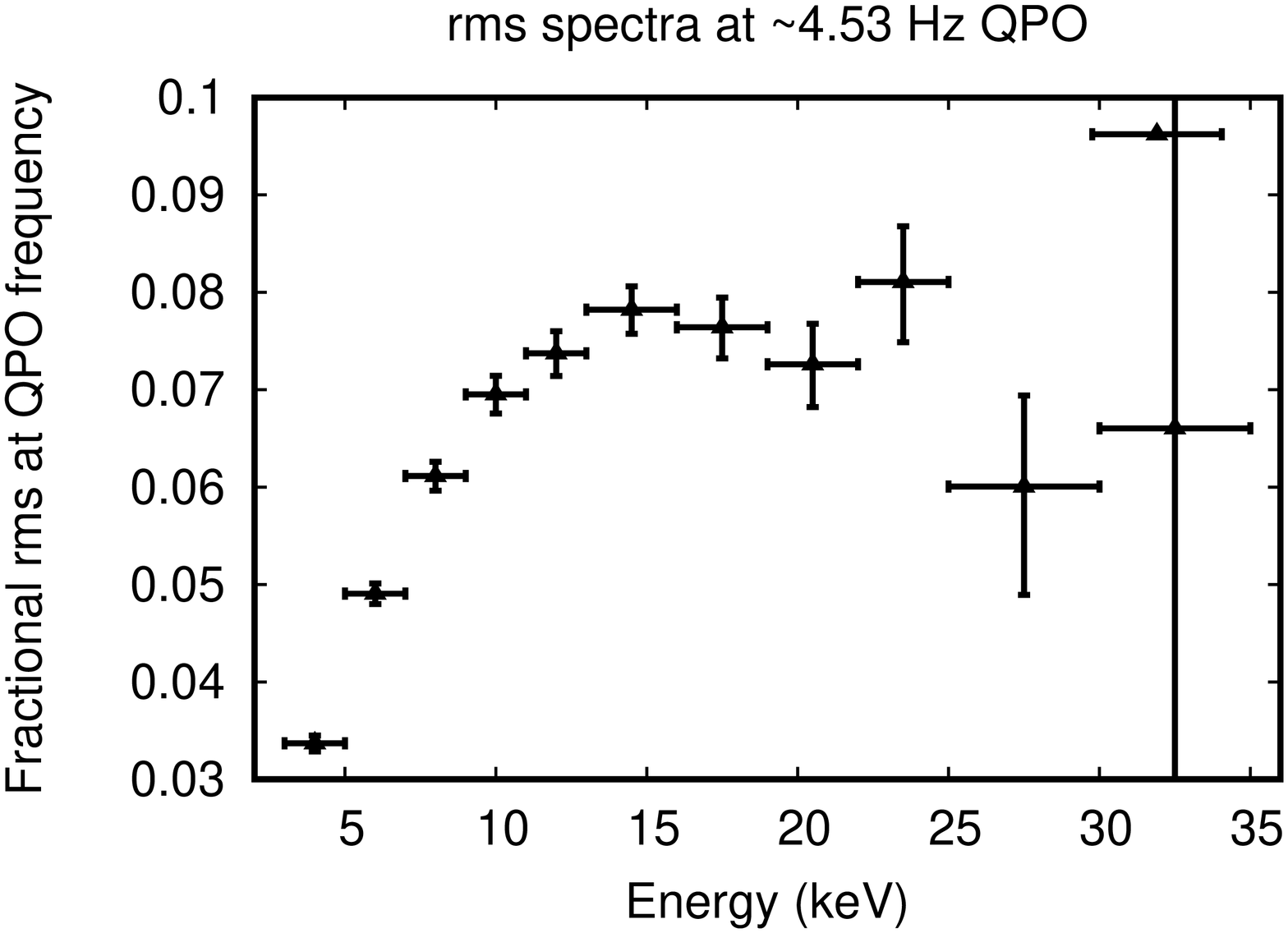}
\centering \includegraphics[width=0.22\textwidth,angle=-90]{fig5g.ps}
\centering \includegraphics[width=0.22\textwidth,angle=-90]{fig5h.ps}
\centering \includegraphics[width=0.22\textwidth,angle=-90]{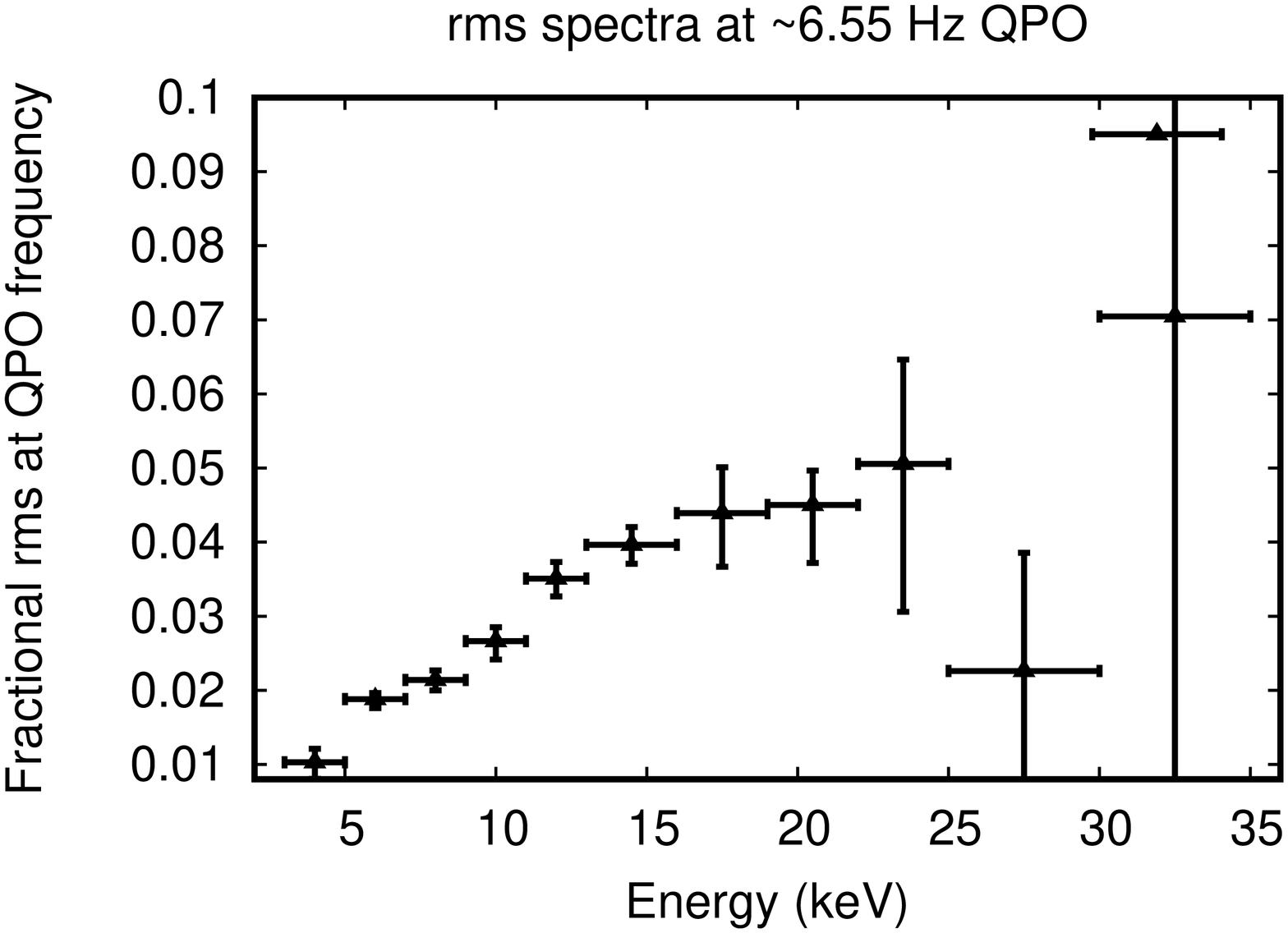}
\caption{In each row, 3.0--80.0 keV fitted power density spectrum in
the frequency range 0.1--10.0 Hz along with its residual (first
column), time-lag spectra at the corresponding fundamental QPO
frequencies (second column) and fractional rms spectra at the same QPO
frequencies (third column) are shown. PDS, time-lag and rms spectra
are calculated at $\sim 2.55$ Hz (top rows), $\sim 4.53$ Hz (middle
rows) and $\sim 6.55$ Hz (bottom rows)respectively. In all lag
spectra, 3.0--4.0 keV band is considered to be the reference band.}
\label{timelag}
\end{figure*}

\begin{figure}
\centering \includegraphics[width=0.27\textwidth,angle=-90]{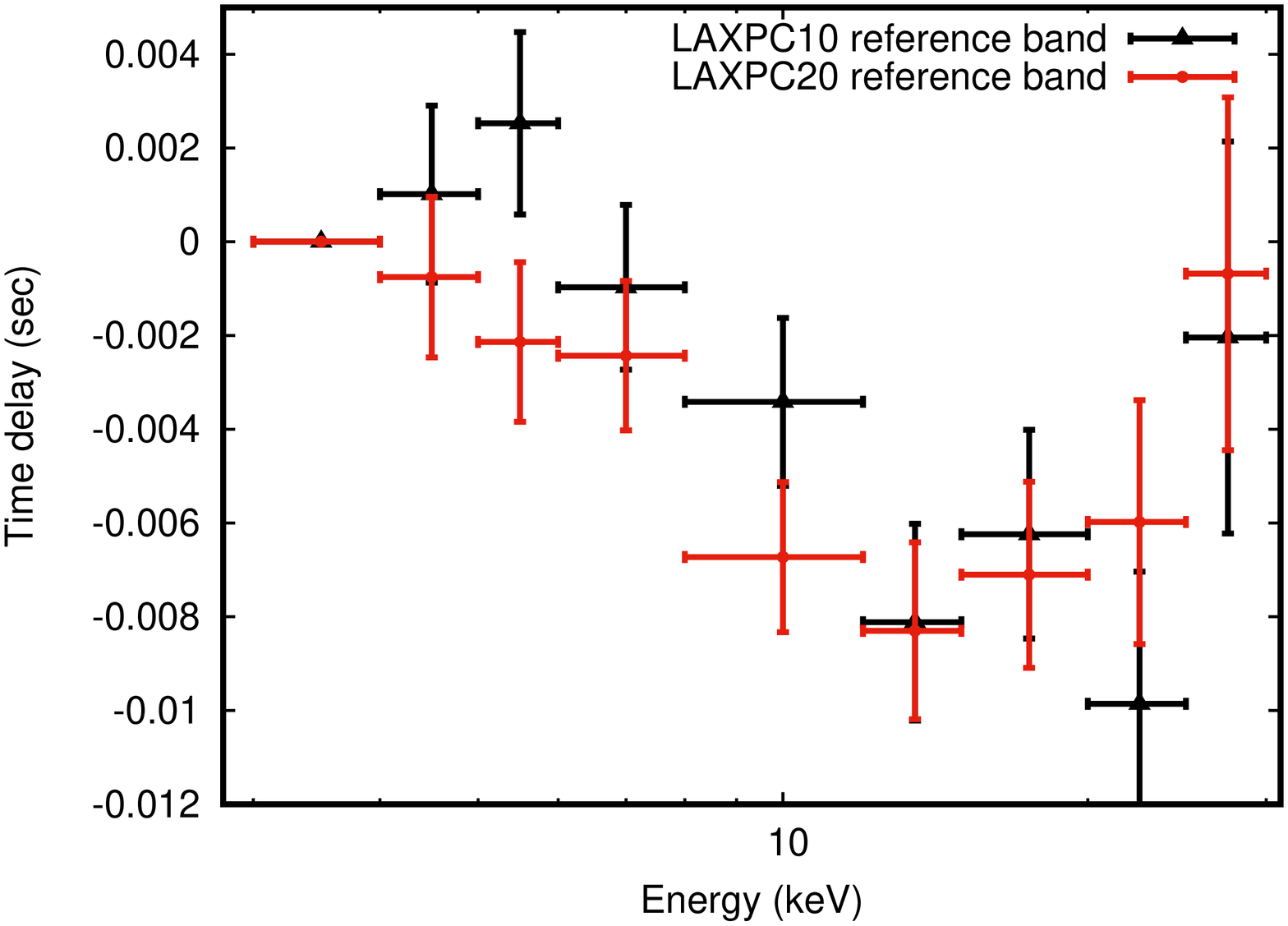}
\caption{Time-lag spectra at $\sim 2.55$ Hz QPO frequencies from two
LAXPC units where lag-spectra of second LAXPC unit ({\tt LAXPC20};
black points) is computed by choosing reference band from first LAXPC
unit ({\tt LAXPC10}) and vice versa (red points). The similarity
between these curves and and for the combined data (top middle panel of Figure 9) shows
that instrumental effects such as dead time is not affecting the
results.}
\label{crosstimelag}
\end{figure}

The LAXPC observations allows the computation of power spectra in
different energy bands. As an illustration, in Figure \ref{power} the
power spectra in three  energy bands --- 3.0--8.0 (left column),
8.0--20.0 (middle column) and 20.0--80.0 (right column) keV are shown
for three representative observations. The dead time ($\sim 42 \mu$sec) corrected Poisson
noise level has been subtracted from each power spectrum. The low
energy power spectra clearly shows a quasi periodic oscillation (QPO)
at 2--7 Hz along with a harmonic for some of the observations which
are fitted using Lorentzians. Complex broad band continuum noise is
seen for all the spectra which has been empirically fitted using a
broken power-law and a few broad band Lorentzians 
depending on the energy. The bottom panels show
residuals in terms of $\chi$ and it is clear that the high quality
data requires more sophisticated analysis rather than the empirical
one adopted here. In particular, the power spectra as well as the QPO
frequencies would not be stationary even during a single observation
and dynamical or flux resolved power spectra are needed to quantify
the changes. While deferring such a detailed  analysis for later, here
we bring out the broad results that the QPO is detected at high
energies and that the shape of the power spectrum significantly
evolves with energy. The shape changes with frequency of the QPO and
there is an additional broad noise component particularly at 8--20 keV
for high QPO frequencies. Such detailed energy dependent power spectra
is a significant improvement over the earlier {\it RXTE} results.  As
expected the QPO frequency varies with flux as shown in Figure
\ref{QPOEdd} which shows the QPO frequency as a function of the
Eddington fraction. It is quite remarkable that the frequency changed
by about 4 Hz with near doubling of the luminosity in a relatively
short time of $\sim$ 5 hours (as shown in Table
\ref{obs}).

\begin{figure*}
\centering \includegraphics[width=0.23\textwidth,angle=-90]{fig7a.ps}
\centering \includegraphics[width=0.23\textwidth,angle=-90]{fig7b.ps}
\centering \includegraphics[width=0.23\textwidth,angle=-90]{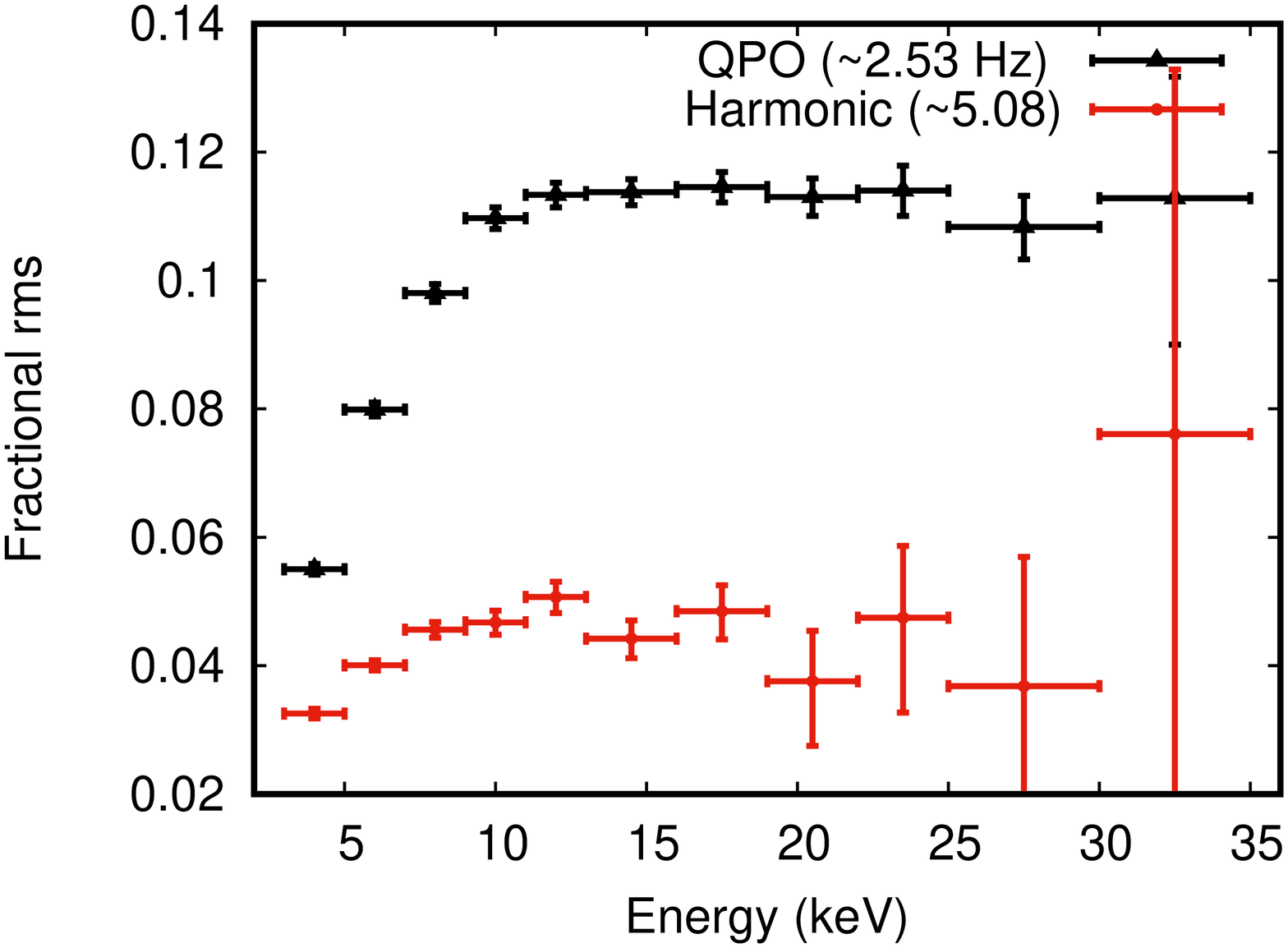}
\caption{Left panel shows energy-dependent time-lag spectra at
$\sim2.55$ Hz fundamental QPO frequency (black circles) and 5.08 Hz
harmonic frequency (red circles) respectively. While soft-lag is
observed at QPO frequency, the lag spectra clearly show hard lag at
harmonic frequency. The middle panel shows phase-lag spectra as a
function of Fourier frequency where the phase-lag is computed between
3.0--5.0 keV and 10.0--20.0 keV energy bands. The position of
$\sim2.55$ Hz QPO and its harmonic at $\sim5.03$ Hz are shown by
vertical dotted lines. Clearly, a phase rotation from negative at QPO
frequency to positive at harmonic frequency is visible. The right
panel shows fractional rms spectra at the fundamental QPO frequency
(black) and its harmonic (red).}
\label{rmsfreq}
\end{figure*}

\subsection{Energy dependent time lag}

The complete event mode data obtained from LAXPCs allows time-lags to
be computed between different user chosen energy bands. This is
demonstrated in Figure \ref{timelag} where the middle column shows the
measured time-lag at the QPO frequency for the three representative
observations. The complete power spectra for 3--80 keV  is shown in
the left column. The time-lag is with reference to the 3--4 keV band
and choice of energy bins is arbitrary. The time lags have a
non-monotonic nature in the sense that typically the photons of $\sim
15$ keV arise before the low energy ones (i.e., a soft lag) but for
higher energies the time lag decreases, i.e., typically the 50 keV
photons have significantly less lag. 
The background corrected root
mean square (r.m.s.) variability of the QPO  increases till
about $\sim 15$ keV and then seems to saturate as shown in the right column 
of Figure \ref{timelag}. However, while the QPO is detected
at high energies and hence the time-lag measurements, 
the 4\% uncertainity in the background rate does not allow for
accurate measurement of the normalized r.m.s. above $\sim 30$ keV. 
This may improve
once more accurate background models are available. While
the energy dependent time-lag  for energies $< 20$ keV is
qualitatively similar to that observed by {\it RXTE}
\citep{qu10,pa13b}, LAXPC reveals that their behavior changes
dramatically for high energies. Time-lag measurements may suffer from
instrumental behavior such as dead time or other such effects which
may produce fake correlations between count variation in different
energy bands. The completely independent electronics of the different
detectors in LAXPC instrument allows one to  check if such effects are
affecting the results. In Figure \ref{crosstimelag} the time-lags for
the $\sim 2.55$ Hz are again measured as a function of energy but here
the reference band (3--4 keV) is taken from {\tt LAXPC10} and the
others from {\tt LAXPC20} (black points) and vice versa (red
points). The behavior of the energy dependent time lag is reproduced
in the cross PCU analysis giving confidence that the overall results
obtained from the combined analysis is correct.

For the QPO observed at $\sim 2.55$ Hz, the next harmonic is detected
clearly in high energy bins and thus one can measure as shown in the
left panel of Figure  \ref{rmsfreq}, the time-lag as function of
energy for both the QPO and its harmonic. As was known before for low
energies using {\it RXTE} data \citep{pa13b}, the time lag for the
harmonic is of the opposite sign (i.e. the time lag is hard) as
compared to the fundamental where the time-lag is soft. The time lags,
$\Delta \tau$, or more conveniently the phase lags $\Delta \phi = 2
\pi {\it f} \Delta \tau$ can also be measured as a function of
frequency $f$ which is shown in the middle panel of Figure
\ref{rmsfreq}. Here the lag is measured between 10--20 keV band and
the 3--5 keV band  where the choice of energy bands is motivated by
the shape of the energy dependent time lag of the QPO. The soft phase
lag at the QPO frequency of 2.55 Hz is clearly seen in  Figure
\ref{rmsfreq} while the phase lag is positive at its harmonic of 5.03
Hz. The right panel of Figure  \ref{rmsfreq} shows the variation of
the fractional rms of the QPO and  the next harmonic.

\section{Summary and Discussion}

In this work, we have presented the first-look analysis of nine orbits
of {\it AstroSat}/LAXPC data of GRS 1915+105, which resulted in nine
observations with exposures varying between 700 sec and 3200 sec. The
primary results are:

\noindent $\bullet$ The energy spectra for each observation can be
well fitted by an empirical model consisting of a disk black body,
thermal Comptonization and an Iron line represented by a broad
Gaussian. The systematic uncertainty of 4\% for the background and
1.5\% for the response provided acceptable fits to all spectra, which
benchmarks LAXPC spectral capabilities.

\noindent $\bullet$ For a single $\sim 3$ ksec observation, the high
frequency power spectrum from 100 -- 50,000 Hz, can be described by a
simple dead time model with an effective dead time of $42.3$
microseconds. Moreover, it is shown that a QPO with quality factor
$\sim 4$ and fractional rms of 5\% would have been detectable easily
till 3000 Hz.

\noindent $\bullet$ Using 0.25 second time bins, hardness intensity
and colour-colour diagrams for a $\sim 3$ ksec observations can be
generated using not only soft but also  hard (e.g., 15--40 keV)
bands. The source shows increasing hardness with intensity in both
soft and hard bands.

\noindent $\bullet$ Power spectra for three energy bins 3--8, 8--20
and 20--80 keV for each observation shows a prominent QPO whose
frequency varies from 2.55 to 6.55 Hz as the flux increases by a
factor of 2 on a 6 hour time-scale. The broad band continuum noise is
complex and shows variation with flux as well as with energy. For
example,` a significant excess noise component appears for higher QPO
frequencies in the 8--20 keV band.

\noindent $\bullet$ At the QPO frequency, the time-lag with respect to
the 3--4 keV band decreases with energy till $\sim 20$ keV as expected
from earlier {\it RXTE} results for the radio quiet $\chi$ class of
GRS 1915+105. LAXPC larger effective area at higher energies reveals
that the lags are non-monotonic and beyond $\sim 20$ keV the time lag
increases with energy.  The time-lag for the next harmonic has 
the opposite sign as the
fundamental and does not seem to roll over at high energies. Since the
electronics of the  PCUs are independent, we have confirmed the
time-lag behavior using one PCU for the reference band and the
another for the different bands.

During the nine observations with LAXPC, we infer that GRS 1915+105
was in the  Radio-quiet $\chi$ class (SPL subclass). There are three
characteristics of these  observations that are consistent with
earlier detection of this class with {\it RXTE}. Namely, (i) the
Radio-quiet $\chi$ class shows a QPO with frequencies
always higher than $\sim2.1$ Hz and the time lag spectra at QPO
frequencies always show a soft lag \citep{pa13b},
(ii) near-simultaneous measurement of Radio flux density using the
Giant Meterwave Radio Telescope (GMRT) at 610 MHz yielded a value of
$\sim6.3$ mJy which is similar to that observed during Radio-quiet
$\chi$ class (see right panel of Figure 1 in \citep{pa13b}) and (iii)
the phase lag between two energy bands as a function of Fourier
frequency is similar to that observed from Radio-quiet $\chi$ class.

These results benchmark the performance of LAXPC as well as bring out
interesting first results which highlight the utility of LAXPC's
larger effective area at high energies and event mode data. For
example, the extension of the variation of the time lag to high
energies brings out the potential to uncover the actual mechanism of
the QPO. The QPO could be due to the precession of the inner disk due
to Lens-Thirring effect and in that case the time-lag may be due to
light travel time effects of the irradiation by the inner disk on the
non-precessing outer one \citep{st98,in09}. The turn over of the
time-lag at high energies would then suggest that perhaps it is due to
the reflection bump being prominent at those energies and hence the
delay with respect to the continuum at $\sim 20$ keV. Detailed
spectroscopic information regarding the contribution of the reflection
component in high energies and the expected light travel time-delays
can reveal whether such a picture is correct or not. Alternatively, in
an entirely different scenario, the QPO could be due to accretion rate
variations which induces a delayed response of the inner disk radius
(DROID). Such a model successfully predicts similar non-monotonic
behavior of the time-lag and rms for the fundamental and next
harmonic for mHz QPO of the same source \citep{mi16}. Again detailed
analysis taking into account the time averaged spectrum need to be
undertaken to find whether such a model is also valid for QPOs with
frequencies of few Hz.

Reanalysis of the data presented with more sophisticated spectral
fitting along with data from the Soft X-ray telescope onboard {\it
AstroSat}, as well as more detailed timing analysis such as dynamic
power spectra, flux and frequency resolved spectroscopy, will provide
unprecedented insight into the temporal behavior of GRS 1915+105 and
other black hole systems.

\section{Acknowledgments}
We acknowledge the strong support from Indian Space Research
Organization (ISRO) in various aspect of instrument building, testing,
software development and mission operation during payload verification
phase. We  acknowledge support of TIFR central workshop during the
design and testing of the payload. We thank the staff of the GMRT that
made radio observations possible. GMRT is run by the National Centre
for Radio Astrophysics of the Tata Institute of Fundamental Research.

﻿

\end{document}